\documentclass[aps,prb,twocolumn,superscriptaddress]{revtex4}
\usepackage{graphicx}
\begin{document}
\title{First principle study of structural stability and electronic structure of CdS nanoclusters}
\author{S. Datta}
\affiliation{S.N. Bose National Centre for Basic Sciences,
Kolkata 700098, India}
\author{ M. Kabir}
\affiliation{S.N. Bose National Centre for Basic Sciences,
Kolkata 700098, India}
\author{ T. Saha-Dasgupta}
\affiliation{S.N. Bose National Centre for Basic Sciences,
Kolkata 700098, India}
\author{ D. D. Sarma}
\affiliation{ Centre for Advanced Materials,
Indian Association for the Cultivation of Science, Jadavpur,
Kolkata 700032, India}
\affiliation{ Solid State and Structural Chemistry Unit, Indian Institute of Science, Bangalore 560012, India}
\date{\today}

\begin{abstract}
Using first-principles density functional calculations, we have studied the structural stability of
stoichiometric as well as non-stoichiometric CdS nanoclusters at ambient pressure with diameters 
ranging up to about 2.5 nm. Our study  reveals
that the relative stability of the two available structures for CdS, namely zinc blende and wurtzite,
depends sensitively on the details like surface geometry and/or surface chemistry. The associated
band gap also exhibits non-monotonic behavior as a function of cluster size. Our findings may 
shed light on reports of experimentally observed structures and associated electronic structures 
of CdS nanoclusters found in the literature.
\end{abstract}
\maketitle
\noindent

\section{Introduction}

The study of physical properties of systems with reduced dimensionality, such as nanoclusters, has
become one of the challenging field of research in recent years. For semiconductor nanoclusters, there
 is a remarkable increase in the band gap compared to the bulk value, as the size of the cluster
 decreases. This has opened up immense technological possibilities in diverse fields such as solar cells\cite{solar},
 electroluminescent devices \cite{device} and possible electronic devices. Among various
 semiconductor nanoclusters, II-VI semiconductor CdS has received a lot of attention, primarily due to
 the following facts, 
(i) CdS is a direct gap semiconductor
with rather large band gap of about 2.5 eV\cite{gap} (ii) quantum confinement effect can be reached quite easily
because of the large excitonic Bohr radius $\approx$ 3 nm \cite{Bohr} (iii) CdS can be synthesized
experimentally rather easily in the size range required for quantum confinement.
The situation, however,
is complicated by the fact that the reduction in particle size seems to influence the structural stability of one
 phase over the other,\cite{new} in a way that is very little understood so far.

Bulk CdS stabilizes in hexagonal wurtzite (WZ) structure. Two other crystal structures, cubic zincblende
 (ZB) and rocksalt structures, in addition to the WZ one or even simultaneous presence of several crystallographic
 phases, have been reported in literature \cite{cubic_results,hexa_results,mixture} for nanoclusters.
 The formation of rocksalt structure is reported only at
 a high pressure,\cite{zbtorsalt} which we exclude from our present discussion, focusing only on the relative 
stability of WZ and
 ZB crystal structures at ambient pressure condition. This is an important issue since not only the band gap
depends on the crystallographic structure of the
nanocrystal,\cite{new} all physical properties such as the effective
masses depend on the underlying crystal structure. However, the
structural similarity between WZ and ZB and the associated
small differences in cohesive energies of the order of few tens of meV/atom make the situation complex.
While such an interesting issue has drawn attention in past and have lead to theoretical analysis
based on parametrized tight-binding models, \cite{springborg1,springborg2} to our knowledge no rigorous first-principles
study exist to address this issue. Very little is also known about the details of the experimental situation,
{\it e.g.} the stability of non-stoichiometric versus stoichiometric clusters, the role of passivator
and their influence on structural stability. In absence of detail knowledge of the experimental
scenario which may also vary in different experimental condition like synthesis route, we considered
in the following the {\it ab-initio} theoretical study of stability of both stoichiometric and non-stoichiometric
clusters, naked as well as passivated. Our study shows that the relative stability between WZ-structured
and ZB-structured clusters are governed by the details of surface geometry and surface chemistry. 
In case of passivated clusters, we have also studied the associated band gap as a function of 
cluster size which depending on specific case also shows highly nonmonotonic behavior.

\section{Building up of clusters}

\begin{figure}
\includegraphics[width=7.5cm,keepaspectratio]{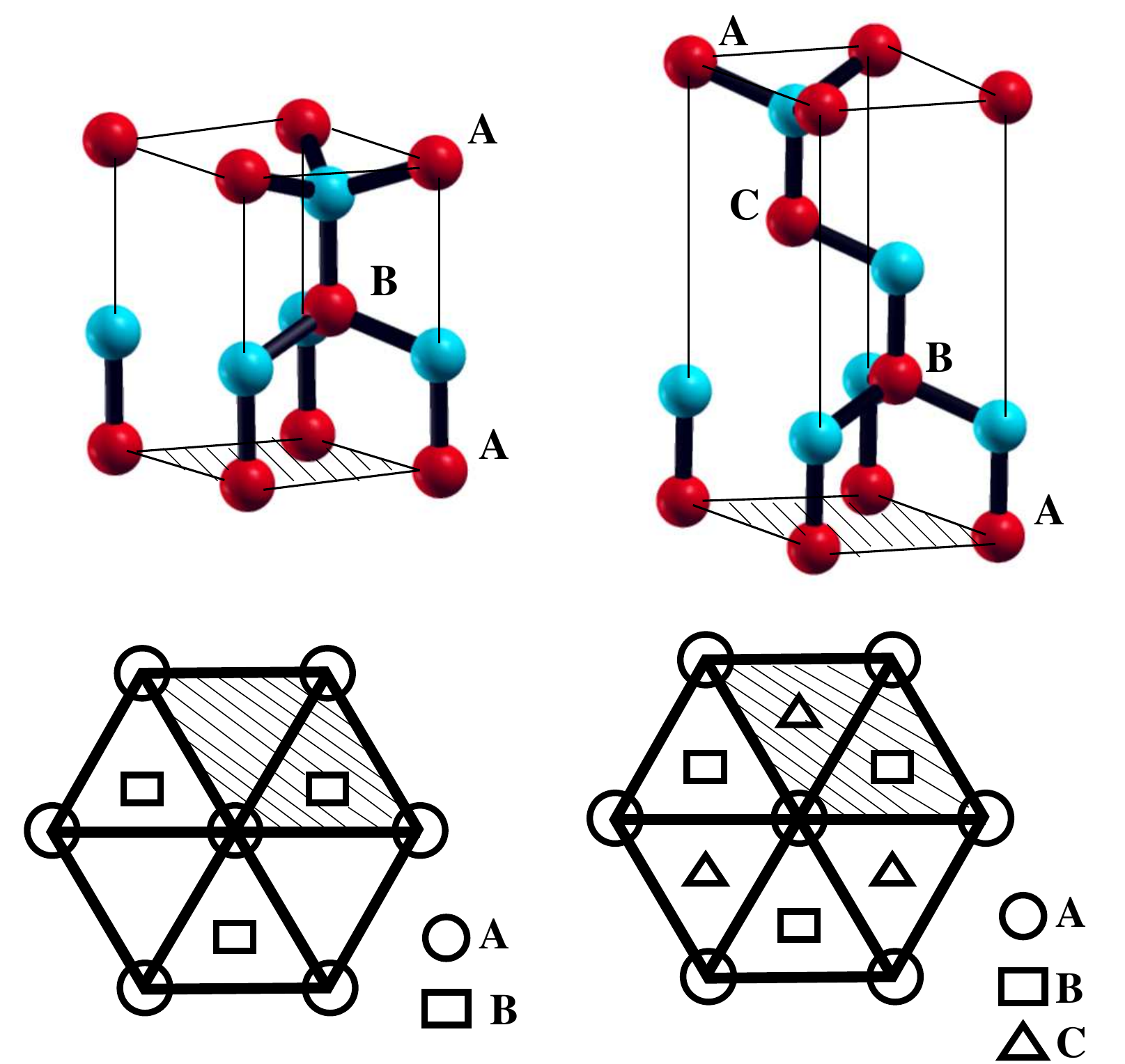}
\caption{(Color online) Crystal structure of CdS in wurtzite (left panel) and zinc blende (right panel)
symmetry. Two differently colored atoms denote Cd and S. Wurtzite structure shows the ABAB$\ldots \ldots$
stacking of atoms along [0001] direction while zinc blende structure shows the ABCABC$\ldots \ldots$ stacking
 of atoms along [111] direction.}
\label{struc}
\end{figure}

The WZ to ZB transformation involves change in symmetry from hexagonal to cubic, while keeping the
nearest neighbor atomic co-ordination fixed at four. The ZB lattice consists of two interpenetrating
face centered cubic lattices of Cd and S, displaced from each other along the body diagonal by
$a_{ZB}/4$, $a_{ZB}$ being the cubic lattice constant. On the other hand, WZ lattice consists of two
interpenetrating hexagonal closed packed lattices, one displaced from another by $3c/16$ along the $c-$axis.
These result into two different stacking sequences: ABAB $\ldots$ along [0001] direction for WZ and ABCABC$\ldots$
along [111] direction for ZB (see Fig.\ref{struc}).

The stable phase also seems to depend on the shape of the
nanocrystal,\cite{nanolett} with majority of the
 experimental results being for spherical nanoclusters. To generate nanocluster of geometry closest to spherical shape,
we have built up the cluster shell by shell. To generate non-stoichiometric clusters, we have taken the 
cluster center on a Cd atom or S atom. In both ZB and WZ structures, each Cd(S) atom is tetrahedrally surrounded by S(Cd)
 atoms in its immediate neighborhood.\cite{note1} Therefore, the surface of a non-stoichiometric cluster, generated
via above-mentioned prescription contains 
only single species of atoms: a non-stoichiometric nanocluster of even number of shells has same kind of atoms at the surface
and at the center, while a non-stoichiometric nanocrystal with an odd number of shells has two dissimilar types of atoms 
at the center
 and at the surface. On the other hand, stoichiometric clusters are generated by putting the center of the sphere
 on the midpoint of Cd-S bond. In this case, each shell contains equal number of Cd atoms and S atoms. Assuming spherical shapes,
the diameters of ZB and WZ clusters of N atoms are given by,
 d$_{ZB}= \left[\frac{3N}{4\pi}\right]^{\frac{1}{3}}a_{ZB}$ and  d$_{WZ}=
\left[\frac{3N}{2\pi}a^2c\right]^{\frac{1}{3}}$ where $a$ and $c$ are the lattice
constants for wurtzite structure in the $ab$ plane and along the $c-$axis. 
In Table \ref{tab:size}, we list the cluster sizes in order of increasing shell numbers for 
both stoichiometric and non-stoichiometric clusters.

\begin{table}
\caption{\label{tab:size} Number of atoms and diameters of both stoichiometric and non-stoichiometric CdS clusters listed in order of increasing shell sizes for wurtzite structure. The corresponding values for the zincblende
 structure are shown within parenthesis.}
{\begin{tabular}{ccccccc}
\hline
\hline
Shell & & \multicolumn{2}{c}{Stoichiometric}&  & \multicolumn{2}{c} {Non-stoichiometric}    \\
 no.  & & N & diameter($\r{A}$)        &  & N    & diameter($\r{A}$)  \\
\hline
1       &  &  8 (8)            & 7.57 (7.21)                 & & 5 (5)                & 6.47 (6.16)  \\
2       &  &  26 (26)          & 11.21 (10.68)               & & 17 (17)              & 9.73 (9.27)  \\
3       &  &  58 (56)          & 14.65 (13.79)               & & 42 (41)              & 13.15 (12.43) \\
4       &  &  114 (110)        & 18.35 (17.27)               & & 86 (83)              & 16.70 (15.72)   \\
5       &  &  192(184)         & 21.83(20.50)                & & 153(147)             & 20.24(19.02)\\
6       &  &  306(294)         & 25.50(23.97)                & & 249(239)             & 23.80(22.37)\\
7       &  &  452(432)         & 29.04(27.25)                & & 379(363)             & 27.38(25.71) \\

\hline
\hline
\end{tabular} }
\end{table}

\section{Computational Details}

We have carried out first principles electronic structure calculation within the 
framework of density functional theory 
for the constructed nanoclusters. We have used projector 
augmented wave (PAW) basis\cite{paw1,paw2} and 
local density approximation (LDA) for the
exchange-correlation functional as implemented in the Vienna
Ab-initio Simulation Package.\cite{vasp} The kinetic energy cut-off of
the plane waves used in the calculations is 
280 eV which gives convergence of total energy sufficient to
discuss the relative stability of various phases. To check the
validity of our calculations, we computed the cohesive energy of
bulk WZ and ZB CdS. Our computed values of -2.653 eV/atom for the
cohesive energy of WZ CdS and 9.7 meV/atom for ZB-WZ energy
difference agree well with published results.\cite{bulk} Finite size
cluster calculations were carried out using the supercell technique
where a finite sized cluster is positioned within a cubic supercell.
The cell dimension is set by the condition that each repeated
cluster in the periodic lattice is separated by a vacuum layer of at
least 12 \AA, large enough so as to avoid the interaction between
the clusters. To check the effect of optimization of geometry in certain 
specific cases, we relaxed the surface atoms keeping
the core of the cluster fixed at ZB or WZ symmetry. This
is a reasonable approach, considering the fact that previous
studies where relaxation has been carried out for the entire
cluster, \cite{springborg1} showed that the structural relaxation
was mostly confined to the surface layer.
Relaxations are performed using conjugate gradient and 
quasi-Newtonian methods until all the force components are less than a threshold value 0.01 eV/\AA.
 The reciprocal space integration in all cases have been carried out with
$\Gamma$ point which is justified by the large dimension of the
cubic supercell.\\
In order to study the role of passivator on the structural stability problem
and the band gap problem we have also considered {\it ab-initio} calculations
in presence of passivators. 
The surface of a naked semiconductor nanoparticle often contains
electronically active states because of unsaturated surface bonds or dangling bond states. Surface
passivation aims to rebond these dangling bonds with some passivating agent while maintaining
the local charge neutrality
of the whole system. In experiment, organic molecules
are often used to passivate nanoclusters.
Owing to the complexities and the numerous degrees of freedom of these 
passivation agents, it is not 
easy to do calculation with such passivators. To mimic the role of passivator in {\it ab-initio} calculation, 
several simpler atomistic models have been proposed. \cite{prepassivation,passivation} We will follow the recipe by 
Chelikowsky {\it et.al.},\cite{passivation}
which is applicable for both stoichiometric as well as non-stoichiometric clusters.
The proposed recipe
requires use of two different kind of fictitious hydrogen atoms, $H^\ast$, to passivate the dangling bonds of CdS
nanoclusters. To keep the passivated cluster
neutral, one species of the fictitious atoms is chosen to have a nuclear charge of
1+$\eta$ and valance electron charge of -(1+$\eta$), where $\eta$ is a positive number. These atoms are
bonded with Cd atoms. The other species of atoms is chosen to have a nuclear charge of 1 - $\eta$ and a valance electron charge
 of -(1-$\eta$). These atoms are bonded to the S atoms. The value of $\eta$ for which gap is maximum in
 the curve of gap vs $\eta$, is found to be 0.5 for II-VI semiconductor nanoparticles.\cite{passivation}
The bond lengths of $H^\ast$-Cd and $H^\ast$-S were determined from two model systems, Cd$H^\ast_4$ and S$H^\ast_4$, in which
bond length are fully optimized. The orientation of $H^\ast$ around Cd and S is fixed in the same tetrahedral 
orientation as it is in ZB or WZ structure.

\section{Results}

\subsection{Energy stability}

\subsubsection{Unpassivated stoichiometric clusters}

\begin{figure}
\includegraphics[width=7.5cm,keepaspectratio]{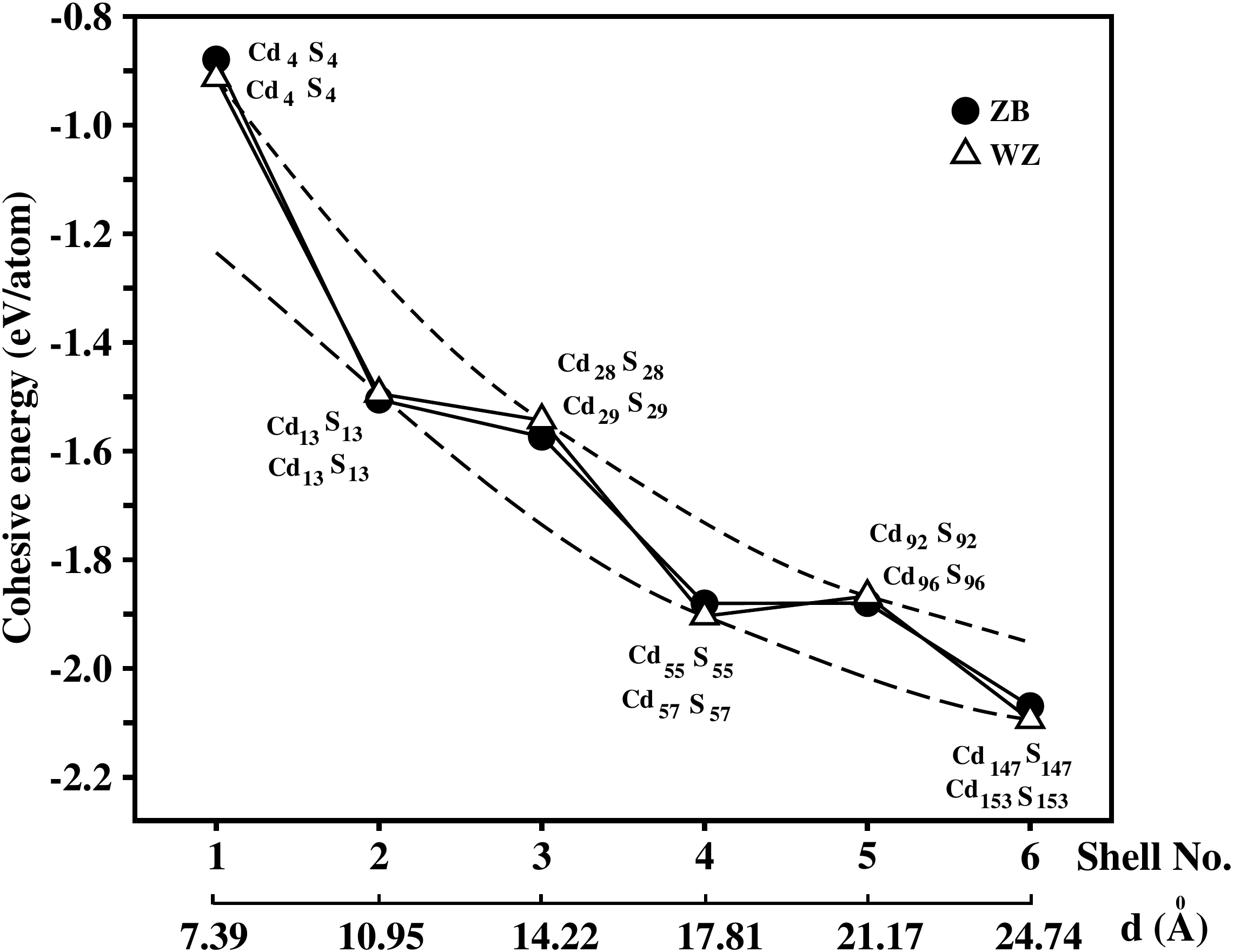}
\caption{Variation of cohesive energy with size for unpassivated stoichiometric clusters
shown by solid dots in ZB structure and by open triangles in WZ structure. For a particular shell number, the diameter of the ZB and WZ structured cluster differ a bit (cf. Table I), therefore the average diameter is shown in the $x$-axis of the plot. Same convention is followed in all the following plots wherever ap
plicable. Chemical formula for each shell is given both 
in ZB (upper) and WZ (lower) structures.}
\label{cohesive_stoi}
\end{figure}

\begin{figure}
\vskip 0.5cm
\includegraphics[width=7.5cm,keepaspectratio]{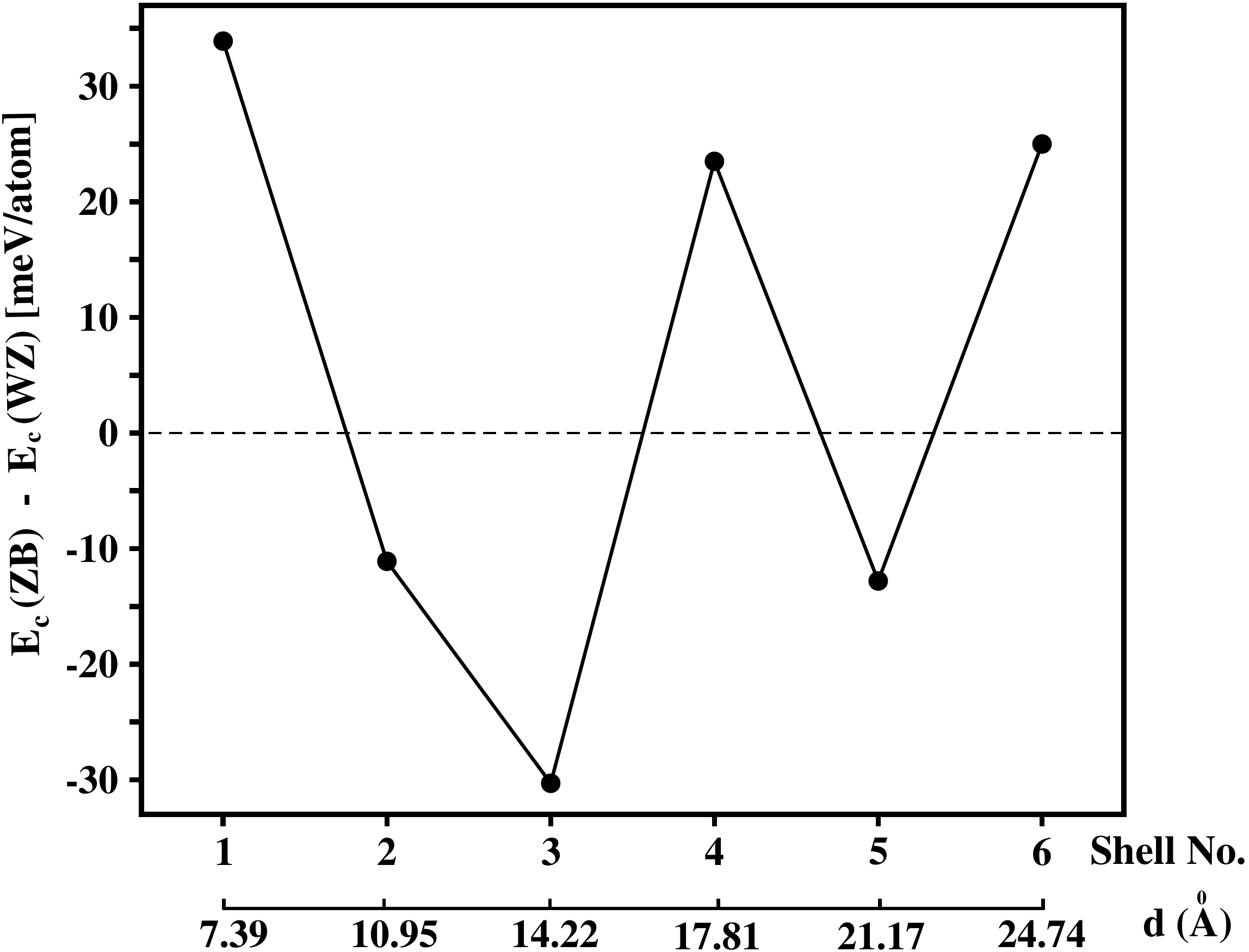}
\caption{Solid dots represent the cohesive energy difference between ZB and WZ structures for each shell size 
in case of unpassivated stoichiometric clusters. Solid line through them is guide to eye.}
\label{cohesive_diff_stoi}
\end{figure}

\begin{figure}
\includegraphics[width=7.5cm,keepaspectratio]{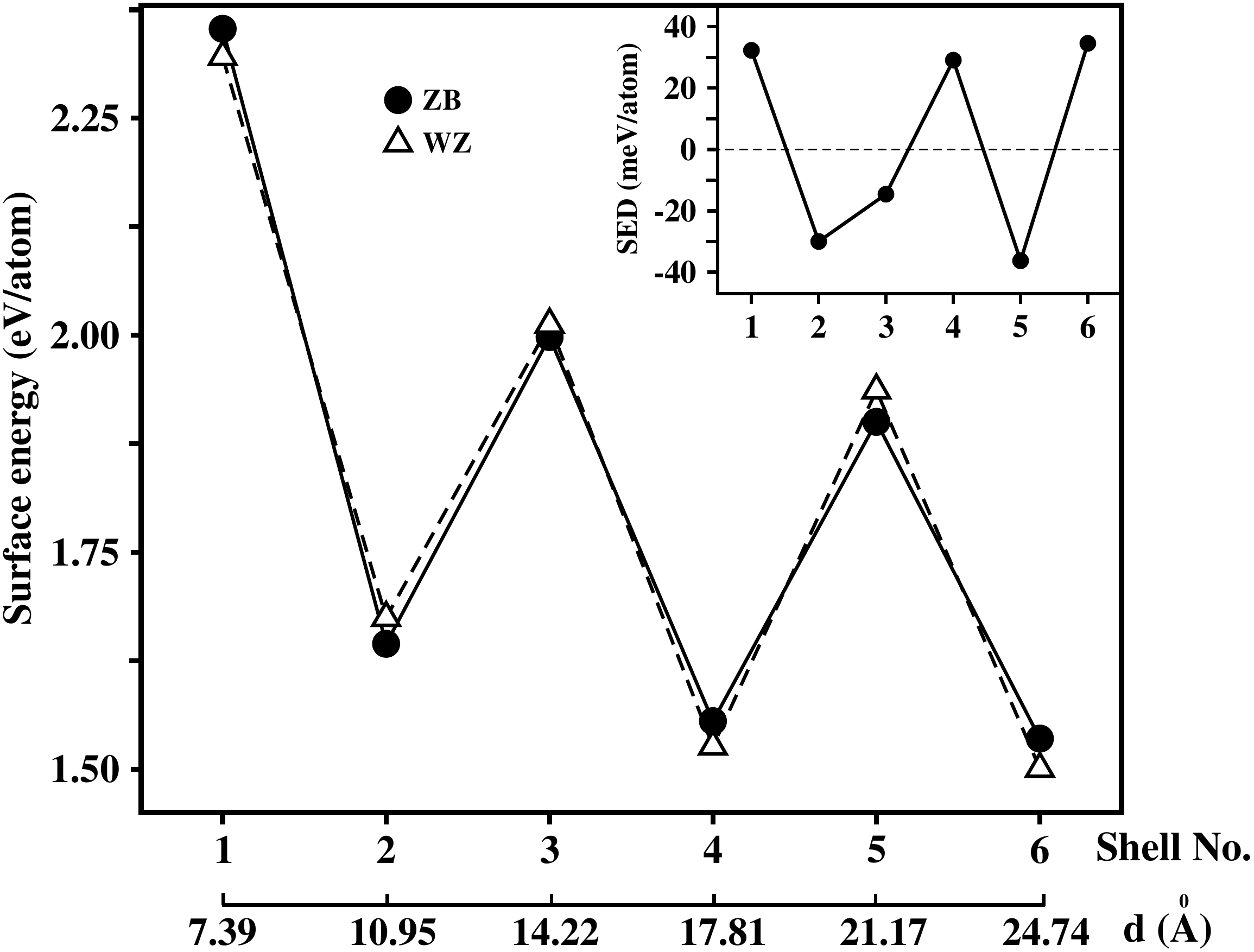}
\caption{Variation of surface energy with cluster size for ZB(solid dots) and 
WZ(open triangles) structures for stoichiometric clusters. The inset shows the 
surface energy difference (SED) between ZB and WZ structures.}
\label{se+sed_stoi}
\end{figure}

Fig.\ref{cohesive_stoi} shows the plot of the computed cohesive energy for unpassivated stoichiometric clusters as a function of growing cluster size, where
we define the cohesive energy per atom as $E_c = \frac{E_{tot}- \sum_{\beta}
n_{\beta}E_{\beta}}
{\sum_{\beta} n_{\beta}}$, $E_{tot}$ being the total energy of the cluster, $E_{\beta}$ being the energy of an
isolated $\beta$ atom ($\beta$ = Cd, S) and $n_{\beta}$ = number of either type of atoms in the cluster. 
The cohesive energy shows an overall decrease with the increase of cluster
size for both the ZB and WZ structures due to the reduction in the ratio of surface atoms to bulk atoms upon increasing cluster size.
However, the variation of cohesive energy is found to be non-monotonic with 
increasing size. Connecting the cohesive energies of all the even shell 
clusters and that of odd shell clusters separately, we find that the even 
shell clusters show higher binding (and therefore lower cohesive energy) than 
the odd shell clusters. Further to study the relative stability between
the ZB and WZ symmetry, we show in Fig.\ref{cohesive_diff_stoi} the cohesive 
energy difference between ZB and WZ structures at each shell size. Whenever the quantity plotted is negative, it implies that the cubic structure is more
 stable, while a positive value signals the hexagonal structure as the stable phase. Our results, as plotted in
 Fig.\ref{cohesive_diff_stoi}, show a general trend that from 3 shell onwards, the odd shell clusters 
 stabilize in cubic ZB structure, while all even shell clusters prefer to form in hexagonal WZ structure.
To check the robustness of our result with respect to structural relaxations
 of the surface atoms, we have also carried out structural relaxation of the surface atoms for 3-shell and 4-shell
stoichiometric nanoclusters, keeping the position of core atoms fixed in ZB or WZ geometry, as explained in section III. 
Although the quantitative values change somewhat (by about 4-10 meV),
 the trend remains the same, {\it i.e.} cluster with odd number of shells like 3-shell stabilizes in 
cubic ZB structure and with even number of shells like 4-shell stabilizes in WZ structure. This is in
 accordance with the finding by Joswig {\it et. al.} \cite{springborg1}
that total energy upon relaxation of such clusters reduces only little.
In order to understand the oscillating stability of the WZ and ZB
structures and also the higher stability of the even shell clusters
compared to odd shell clusters in general, we have analyzed the different contributions to the
total energy. For finite sized cluster, surface effect is important and the 
surface energy contribution to the total free energy plays the dominant role in 
 determining the stable phases. In Fig.\ref{se+sed_stoi} we show the surface energy variation
 as a function of the cluster size, where the surface energy per atom is 
defined as, $\frac{E^{clus} - N \epsilon^{bulk}}{N_s}$, $E^{clus}$ being 
the total cohesive energy of the cluster, $N$ being the total number of atoms in the cluster,
$N_s$ being the number of surface atoms, and $\epsilon^{bulk}$ being the
bulk cohesive energy per atom. We find that even shell clusters have lower surface energy
 and hence more binding compared to odd shell clusters. The inset of Fig.\ref{se+sed_stoi}
 shows the surface energy difference between ZB and WZ structures for each shell size. The 
positive (negative) value of surface energy difference means ZB structure has higher(lower) 
surface energy contribution than 
WZ structure and therefore ZB structure is less(more) stable. This trend in surface energy variation
and that of its difference is in accordance with the 
trend found in the cohesive energy and its difference. 

Furthermore, to understand the nonmonotonic behavior in surface energy or cohesive energy as a function
of cluster size, we calculated the average number of dangling bonds per surface atom. The number of 
dangling bonds are the total number of unsaturated bonds, defined as $\sum_{i} (4 -z_{i})$, where $z_i$ 
is the coordination of $i$-th
surface atom and the summation over $i$ involves summation over all surface atoms for a particular
cluster size. We find that the average
number of dangling bonds are larger for odd shell clusters, giving rise to increased surface states
and hence larger surface energy contribution, thereby explaining the nonmonotonic behavior of cohesive
energy. While nonmonotonic behavior of cohesive energy has been reported in theoretical 
calculations in past \cite{springborg1}, such systematic behavior and its analysis by application of accurate 
first-principles calculations, to our knowledge has not been demonstrated before. The analysis in
terms of dangling bonds not only explains the higher stability of the even shelled clusters, but
also explains the relative stability of WZ and ZB structured clusters since the difference
 of average number of dangling bonds per surface atom between ZB and WZ structured clusters
oscillates between positive and negative values, being positive for even shelled clusters and
negative for odd shelled clusters. The difference is zero for 1 and 2-shelled clusters since
the local co-ordination of immediate neighbors is identical between ZB and WZ clusters and
the difference shows up only beyond 2nd nearest neighbor.

\begin{figure}
\includegraphics[width=7.5cm,keepaspectratio]{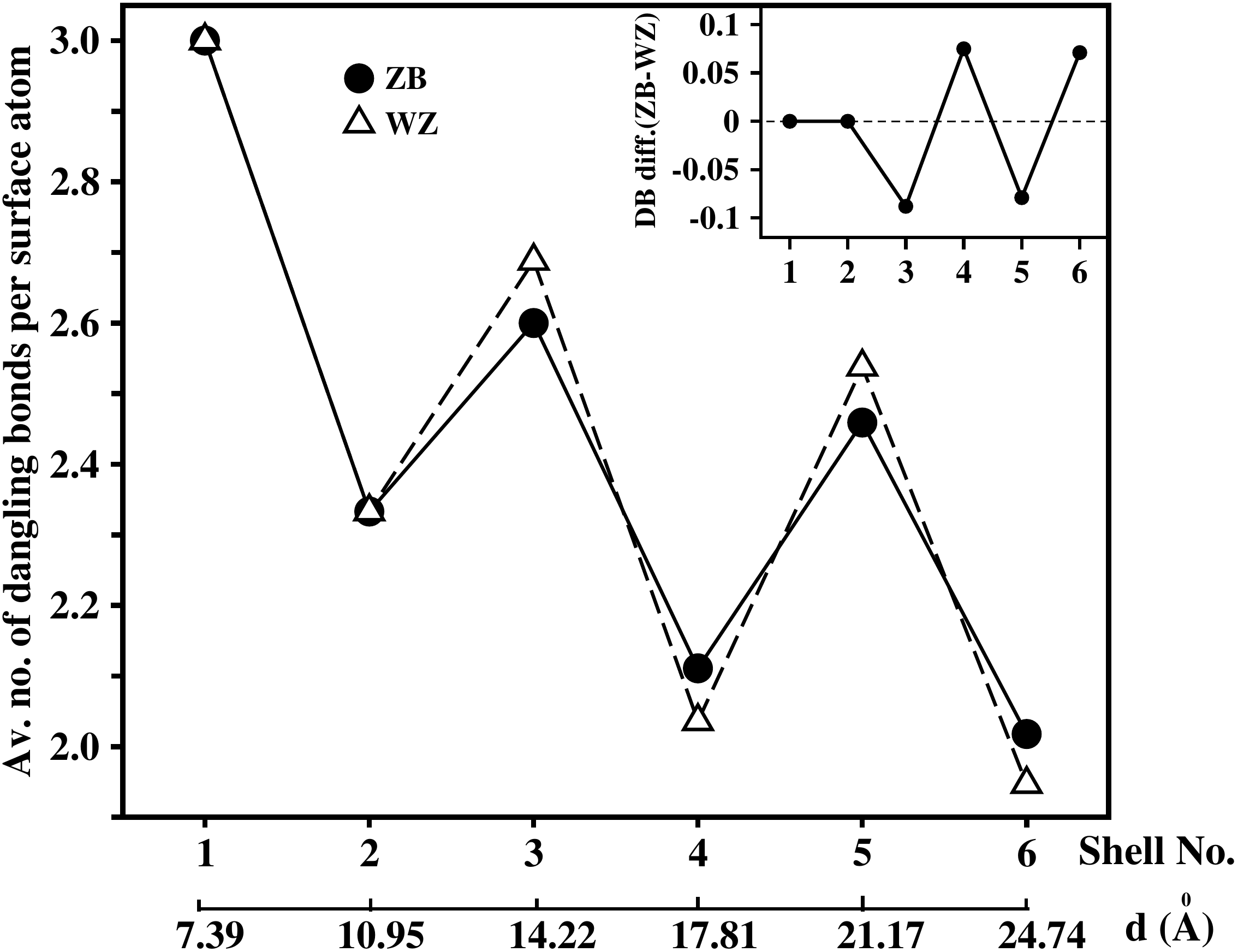}
\caption{Average number of dangling bonds (DB) per surface atom for unpassivated stoichiometric 
clusters in ZB (solid dots) and WZ (open triangles) structures. The inset shows difference between ZB and WZ 
structures.}
\label{db+dbdiff_stoi}
\end{figure}

\subsubsection{Unpassivated non-stoichiometric clusters}

\begin{figure}
\includegraphics[width=7.5cm,keepaspectratio]{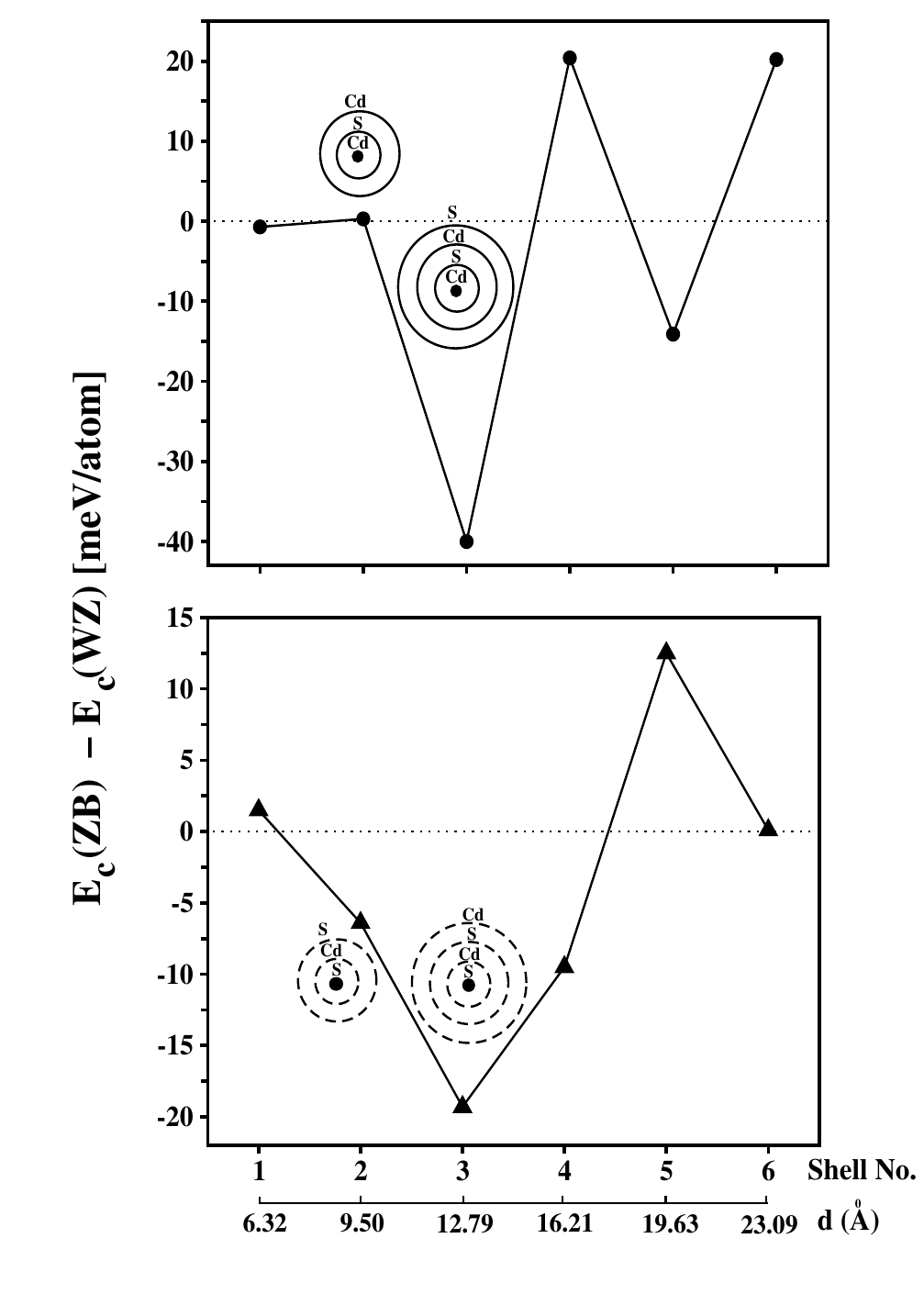}
\caption{Cohesive energy difference between ZB and WZ structures for each shell size 
in case of unpassivated non-stoichiometric clusters. Solid dots in the upper panel represent
 the results for Cd-centered clusters while the solid triangles in the lower panel correspond
 to results for S-centered clusters. Note, by construction even shell Cd(S)-centered cluster 
 is Cd(S) terminated and odd shell Cd(S)-centered cluster is S(Cd) terminated (see text).}
\label{nonstoi}
\end{figure}

\begin{figure}
\includegraphics[width=9.0cm,keepaspectratio]{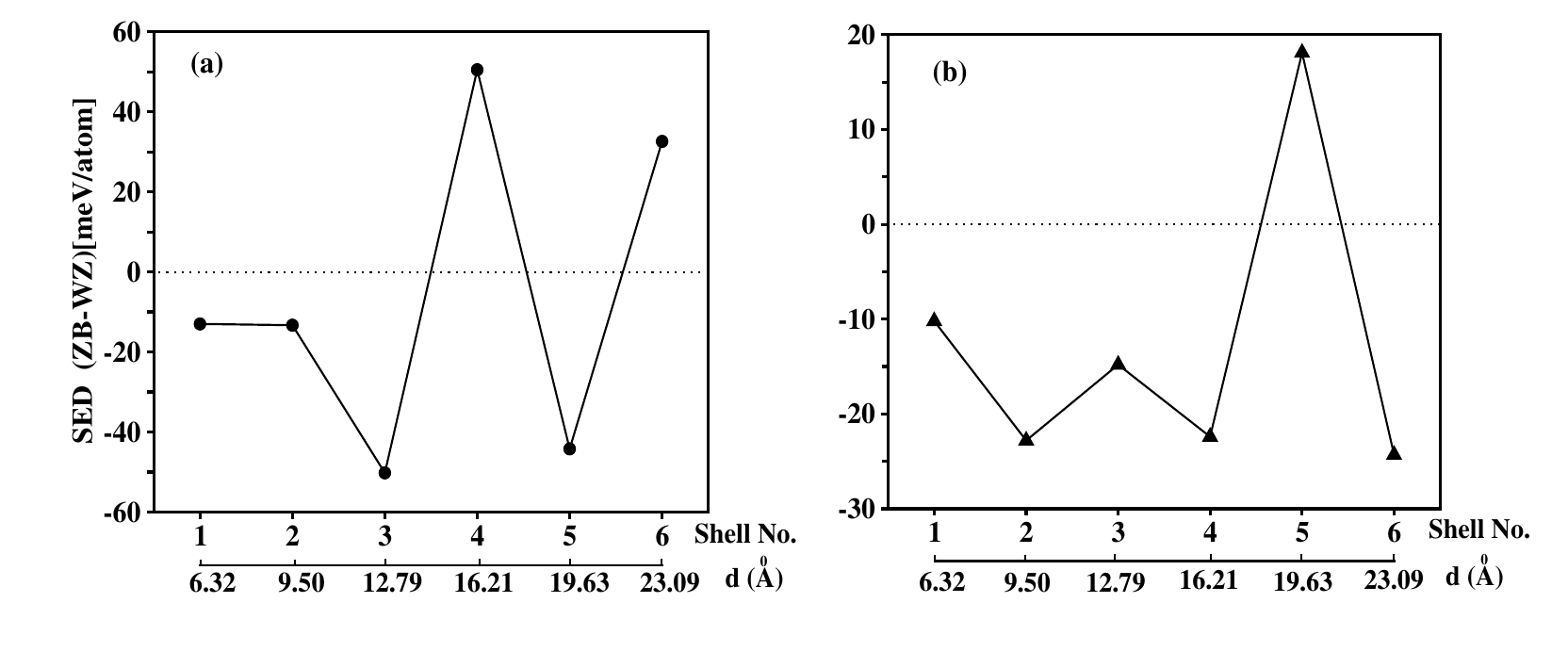}
\caption{Variation of surface energy difference (SED) between ZB and WZ structures, with increasing shell sizes in Cd-centered (representerd by solid dots in (a)) and S-centered (represented by 
solid triangles in (b)) non-stoichiometric clusters. }
\label{SE-non}
\end{figure}

\begin{figure}
\includegraphics[width=7.5cm,keepaspectratio]{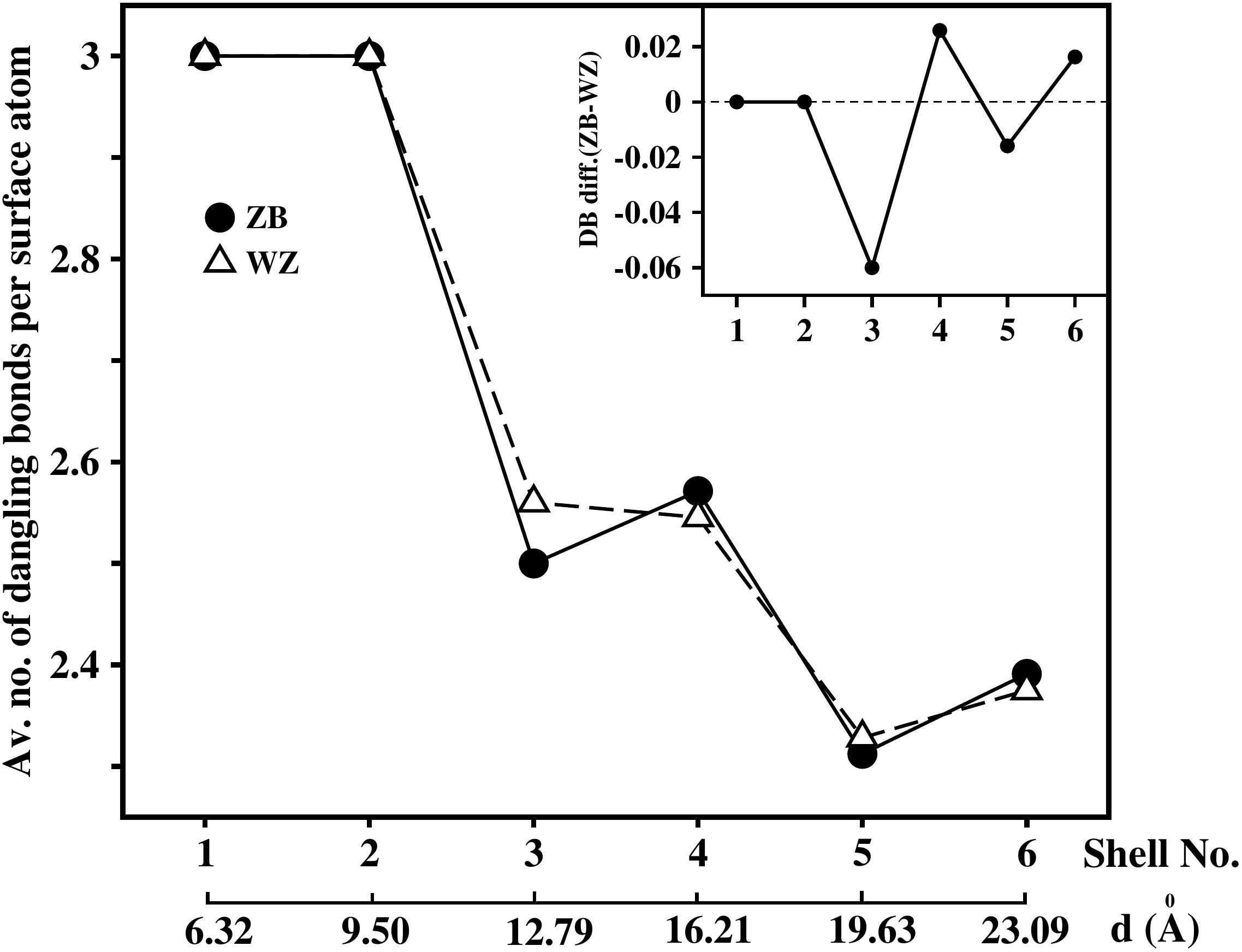}
\caption{Average number of dangling bonds per surface atom (DB) for non-stoichiometric CdS 
clusters in ZB and WZ structures. The inset shows the difference between ZB and WZ structures. Plot is independent of whether the clusters are Cd-centered or 
S-centered. }
\label{db-non}
\end{figure}

Most experimental condition favors synthesize of clusters of non-stoichiometric nature.
It is therefore important to consider clusters having non-stoichiometric composition.
This however leads to complication due to the fact that different cluster sizes have very
different Cd to S ratio, making the comparison of corresponding total energies a difficult
task for which no obvious way exists. However, for a given shell, the Cd to S ratio between
ZB and WZ structured clusters remains almost same\cite{footnote} making the comparison
of their cohesive energy meaningful. In Fig.\ref{nonstoi} we 
plot the difference in cohesive energy between the ZB and WZ structured non-stoichiometric
clusters as a function of increasing cluster size. The plot in the upper panel
exhibits the cohesive energy differences for Cd centered clusters. The plot in the
lower panel exhibits the same but for S centered clusters. We note that unlike the 
case of stoichiometric clusters where the surface composition always consists of
equal number of S and Cd, for non-stoichiometric clusters the surface is formed
exclusively by either S or Cd atoms. For Cd-centered clusters, the even(odd) shell
clusters are Cd (S) terminated, while it is reversed for S-centered clusters. 
From the plot in Fig.\ref{nonstoi}, we find that
nonmonotonic behavior of the relative stability between ZB and WZ phase
persists, on top it shows additional interesting aspect in the sense that whether
an even or odd shell cluster is formed in WZ or ZB symmetry depends on the
terminating layer. Focusing on clusters with shell numbers 4 and 5 in upper pannel
 of Fig.\ref{nonstoi}, 4 shell cluster is found to be WZ structured and 5 shell 
clusters is found to be ZB structured for Cd-centered cluster while moving to the
 lower panel of Fig.\ref{nonstoi}, the reverse trend is found for the S-terminated clusters.
  We note that a 4 (5) shell cluster is Cd (S) terminated in the former cases and S (Cd) 
terminated in the latter cases. As expected this is driven by the oscillating behavior
of the surface energy difference between ZB and WZ structures (shown in Fig.\ref{SE-non}), 
which shows the similar trend as observed in case of the cohesive energy difference.
However unlike in case of stoichiometric clusters, this trend is not explained by the difference 
in the average number of dangling bonds per surface atom between ZB and WZ structured clusters
 (shown in Fig.\ref{db-non}). The difference in average number of dangling bonds predicts 
that the surface energy of a 4-shell ZB structured cluster to be higher than
 that of a WZ structured cluster, hence a 4 shell cluster must form in WZ structure which 
is indeed the case for 4 shell Cd-centered cluster but not for 4 shell S-centered cluster.
 We therefore conclude that an additional effect is operative in case of non-stoichiometric
cluster, namely the surface chemistry effect. The surface chemistry effect
adds on the surface geometry effect in case of Cd centered clusters, while
it acts in an opposite way to that of surface geometry effect in case of
S centered clusters, thereby reversing the trend in the sense odd shell clusters are 
now stabilized in WZ structure and even shell clusters are stabilized in ZB structure.
 The relative stability between ZB and WZ structures in case of non-stoichiometric clusters, 
is driven dominantly by the surface chemistry rather than
the surface geometry as has been found in case of stoichiometric clusters. 

\begin{figure}
\includegraphics[width=8.5cm,keepaspectratio]{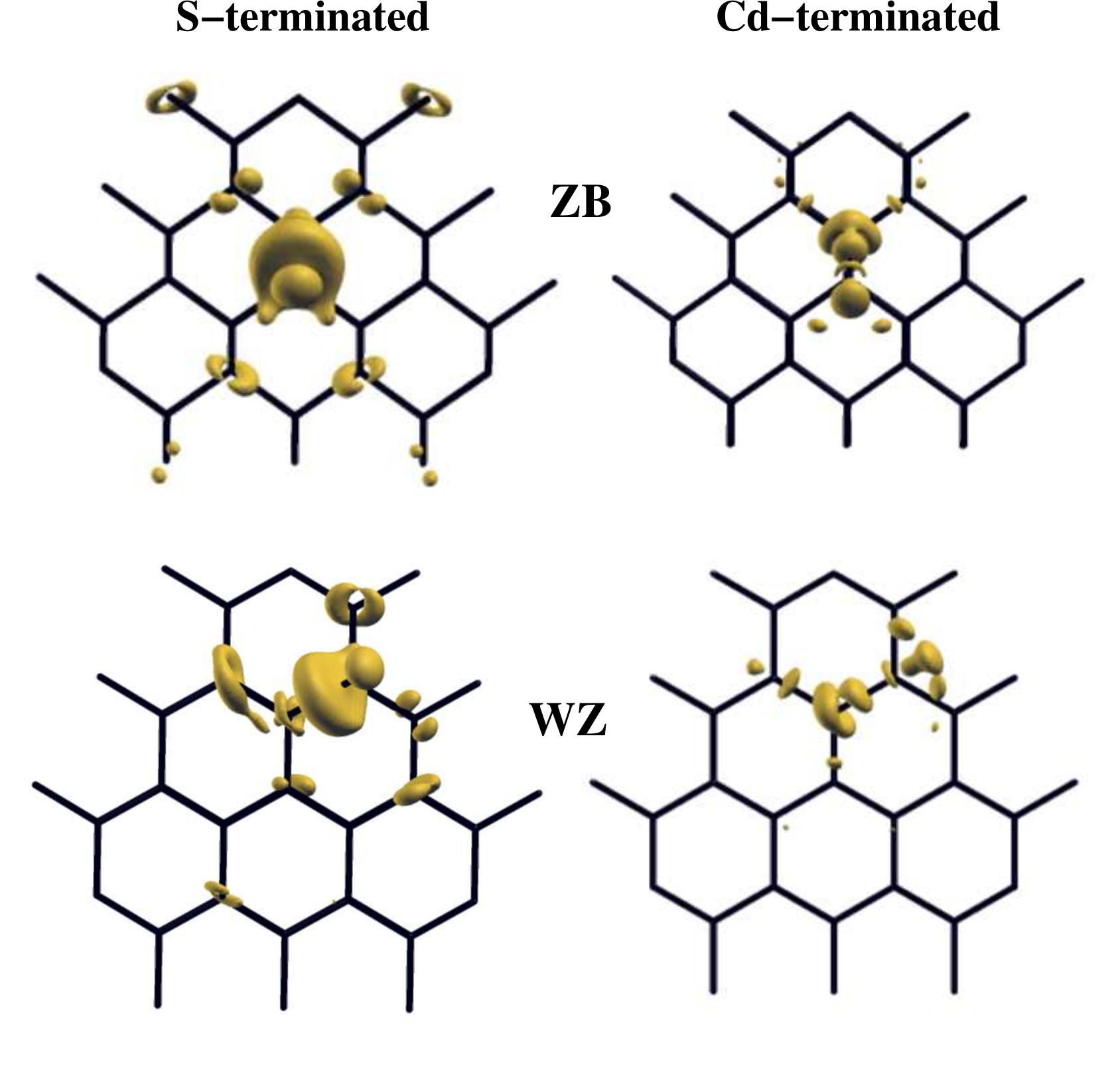}
\caption{(Color online) Charge density contribution ($\delta\rho = \rho_t - \rho_a - \rho_s$) around 
a surface atom of 4-shell non-stoichiometric CdS nanoclusters. The isosurface is chosen 
at .007 e$^{-}$ /($\AA^{3}$).}
\label{charge}
\end{figure}

In order to investigate the microscopic reasons associated with
the surface that drives this effect, we have computed the average charge 
enclosed within a sphere around a Cd atom and that around an S atom within 
a given cluster. There is no unique way to divide space in an AB compound
into A and B regions. Therefore, two choices of sphere radius have
been made: in one case, the spheres were taken to be equal sized
with radius as half the Cd-S bondlength, in another case the choice
of sphere radius was guided by the Hartree plot of the potential. We
carried out calculations for Cd-terminated nanoclusters and
S-terminated nanoclusters as well as for that of the bulk with both
ZB and WZ structures in every case. The following results emerge
from these calculations independent of the structure type. While the
Cd-terminated clusters show similar charge distributions as those of
bulk, S-terminated clusters show about 0.2 fraction of less
electronic charge enclosed within the spheres. This is found to be
true for both choices of sphere radii. This, in turn, would indicate
covalency to be stronger in case of S-terminated clusters resulting
into significant amount of charge residing in the interstitial
region between  Cd and S-centered spheres. This is illustrated in
Fig. \ref{charge}, which shows the charge density distribution
($\delta\rho$) around a surface S atom, and that around a surface Cd
atom for a 4-shell non-stoichiometric cluster, after subtracting the charge 
density of the isolated atom ($\rho_a$) and that
 of the system without the chosen atom ($\rho_s$) from the actual system ($\rho_t$).
 While Cd terminated cluster shows hardly
any change in the region of Cd-S bond,
 there is a significant accumulation of charge around the Cd-S bond 
in case of S-terminated cluster. This provides
a clear evidence for an enhanced covalency and, therefore, reduced ionicity 
in case of S-terminated cluster resulting
into increased stability of ZB phase over the WZ phase. In this context, it is interesting to note that the stability
 of the two competing crystal phases, namely ZB and WZ, changes systematically 
for the bulk systems, CdS, CdSe and
CdTe. While CdS has the WZ structure, CdTe is known to have the ZB form which 
has indeed been explained \cite{soumendu,zunger}
 in terms of increased covalency. This is consistent with the present observation of ZB structure for CdS nanocrystal being stabilized in presence of a S-terminating layer which is also substantially more covalent. 



\subsubsection{Passivated stoichiometric clusters}

In realistic situation, the clusters are grown in presence of some passivating agent. 
Although ideally the role of the passivator is to restrict the growth of the cluster by
saturating the unpassivated dangling bonds, which opens up a clear gap in the
energy spectrum without supposedly changing the intrinsic properties of the
clusters, it may also influence the energy stability of the cluster itself.
This is however very complicated and rather unexplored issue due to the complexity
of various passivating agents used in experiments. A good understanding of the
atomic structure of such complex passivating agents like trictylphosphine (TOP)
or trictylphosphine oxide (TOPO) in many cases is unavailable and it is almost
impossible to deal with such large complexes within an accurate first-principles approach. Very
often, therefore fictitious H atoms are used in theoretical calculations 
for the purpose of passivation. In absence of any other well-defined procedure
we have therefore considered the passivation by fictitious H atoms and in the
following have studied the effect of the passivation on both stoichiometric
and non-stoichiometric clusters.

\begin{figure}
\includegraphics[width=8.5cm,keepaspectratio]{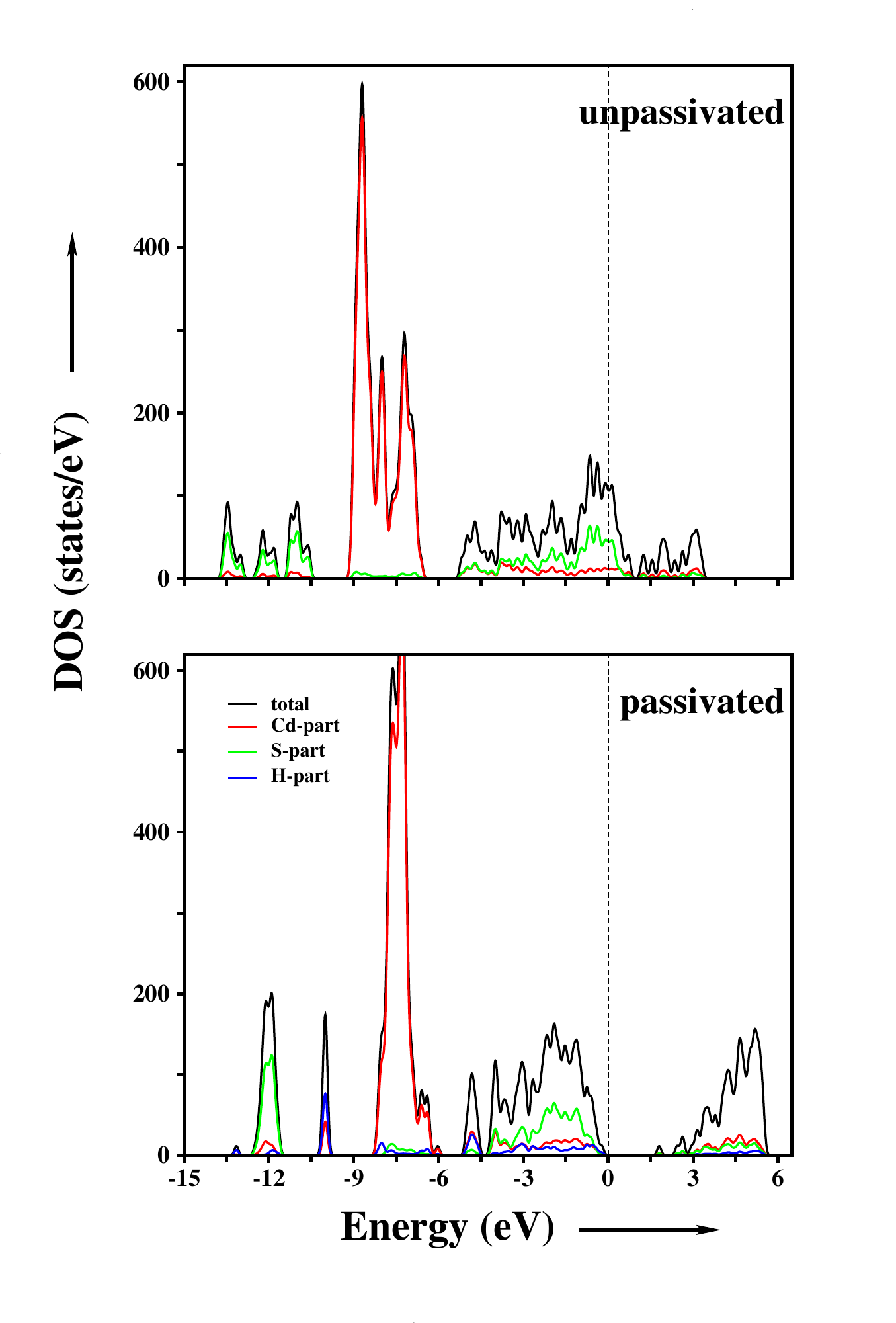}
\caption{(Color online) Density of states for 4 shell ZB structured stoichiometric cluster. 
Upper panel shows the DOS for unpassivated cluster while the 
bottom panels shows the DOS for passivated cluster. Black, red, green and blue lines
correspond to total DOS, Cd-contribution, S-contribution and
that of H-contribution (in case of passivated cluster only) respectively.  }
\label{dos_stoi}
\end{figure}

\begin{figure}
\includegraphics[width=7.5cm,keepaspectratio]{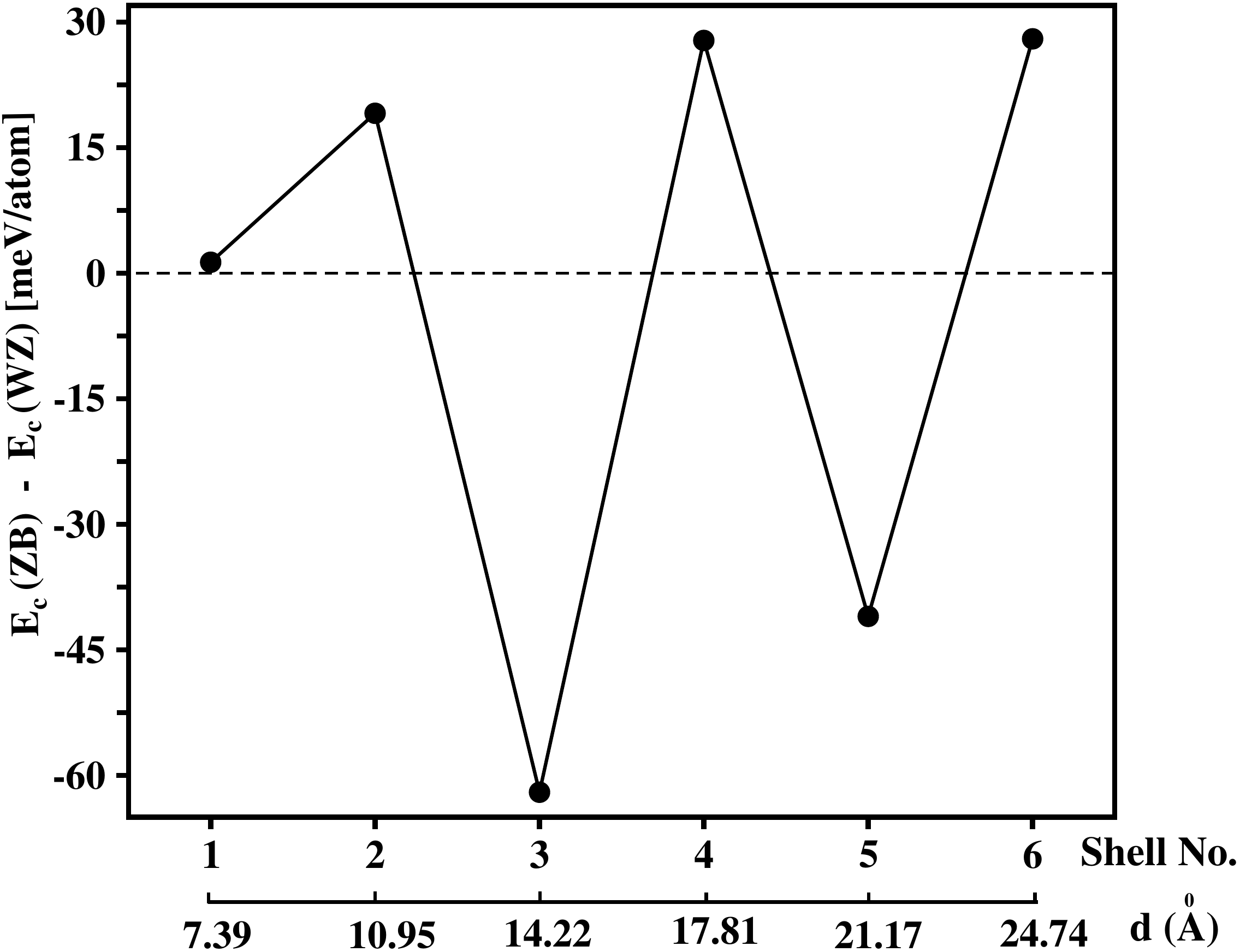}
\caption{Variation of cohesive energy difference between ZB and WZ structures 
with size for passivated stoichiometric clusters.}
\label{cohesive_pass_stoi}
\end{figure}

In Fig.\ref{dos_stoi}, we show the computed density of states of a representative 
stoichiometric CdS cluster with 4 shell and ZB structure in absence and presence
of passivation. We note that assumed passivation could successfully remove the 
states close to Fermi energy, opening up a gap of about 1.5 eV. Similar results
are obtained for clusters with other different shell structures and also with WZ
symmetry. Having been convinced about the proper functioning of the passivator, in
Fig.\ref{cohesive_pass_stoi} we show the variation of the cohesive energy differences
between the ZB and WZ structured stoichiometric CdS nanoclusters as a function of
increasing cluster size. Interestingly we note, that the oscillating behavior
of the relative stability between ZB and WZ structured clusters observed for
naked stoichiometric clusters survives even in presence of passivation, in the
sense the even shelled clusters favor the WZ structure and the odd shell clusters
favor the ZB structure. This presumably is driven by the fact that the difference
in number of bonds with fictitious H atoms between WZ and ZB structures oscillates
as a function of increasing cluster size. As already stated, we have followed in the 
above a simplified treatment of passivating agent, the situation in presence of
realistic passivators need to be explored, which however is beyond the scope
of our present study.

\subsubsection{Passivated non-stoichiometric clusters}

\begin{figure}
\includegraphics[width=8.5cm,keepaspectratio]{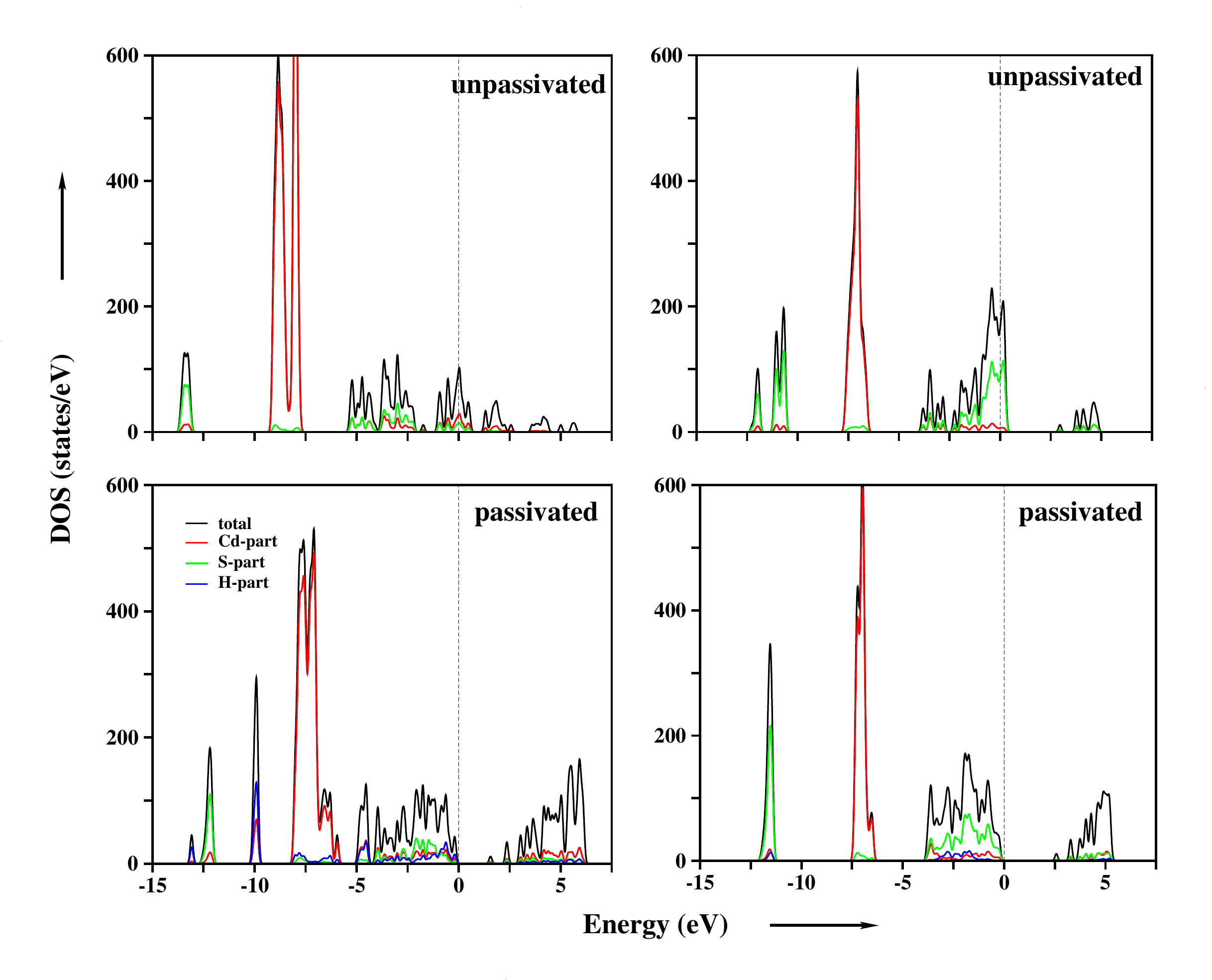}
\caption{(Color online) DOS for 4 shell Cd-centered (left panels) and S-centered (right panels) 
ZB non-stoichiometric cluster. Upper panels correspond to unpassivated case while bottom 
panels correspond to passivated case. Black, red, green and blue lines
correspond to total DOS, Cd-contribution, S-contribution and
that of H-contribution (in case of passivated cluster only) respectively.  }
\label{dos-non}
\end{figure}

\begin{figure}
\includegraphics[width=7.5cm,keepaspectratio]{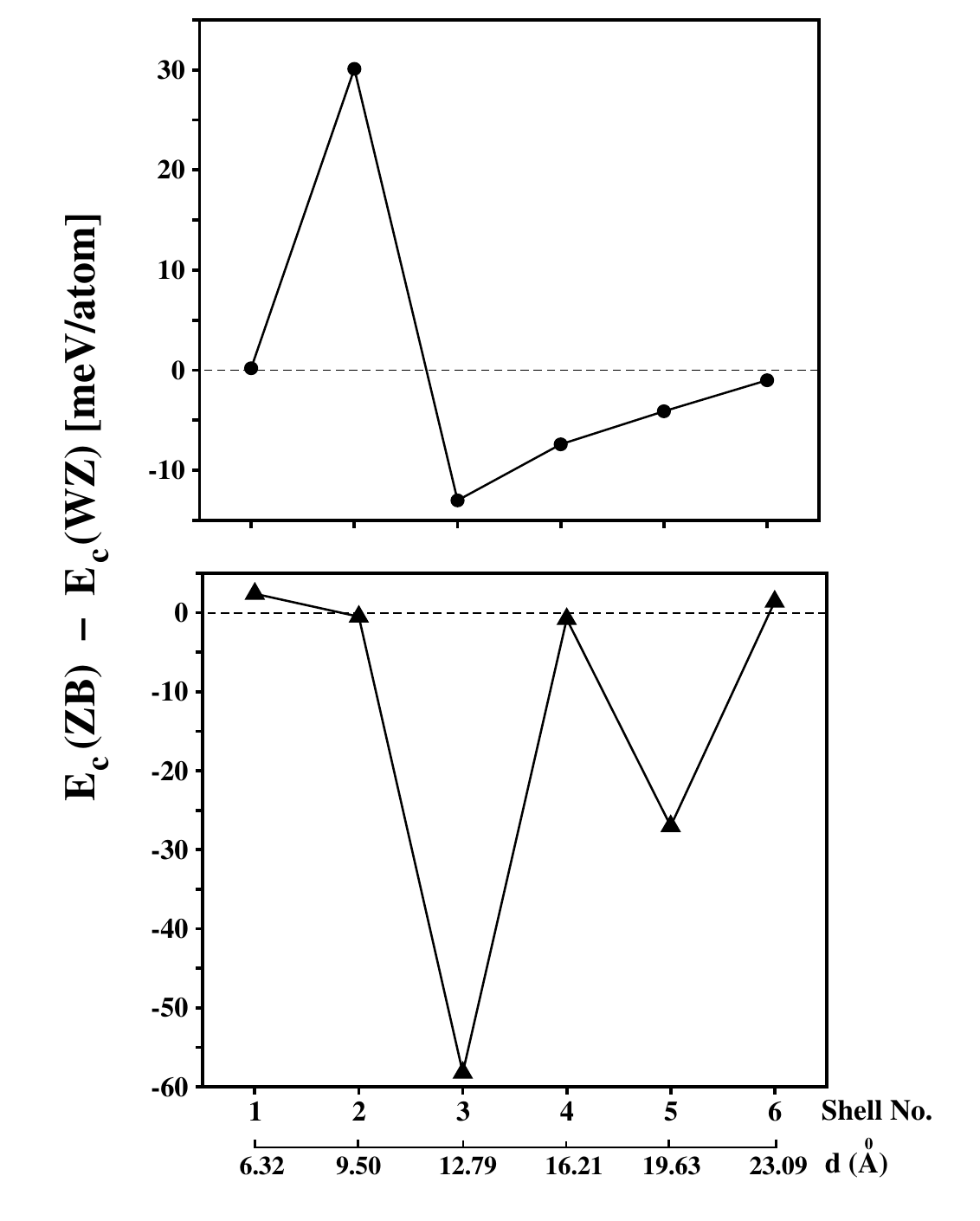}
\caption{Variation of cohesive energy difference between ZB and WZ structures 
with size for passivated Cd-centered (upper panel) and S-centered (lower panel) 
non-stoichiometric clusters.}
\label{cohesive-pass-non}
\end{figure}

In this section, we focus on non-stoichiometric clusters and the role of
passivation in this class of clusters. The passivation has been done following
the same prescription as in case of stoichiometric clusters. Fig.\ref{dos-non}
shows the density of states plot for a representative non-stoichiometric
cluster in absence and presence of passivator. As found in case of stoichiometric clusters, the passivator 
removes the surface states appearing close to Fermi energy in the
unpassivated case and shifts them away from Fermi energy, thereby opening up a clear
gap at the Fermi energy. The corresponding variation in the relative stability
of the ZB and WZ structures for the passivated non-stoichiometric clusters are shown
in Fig.\ref{cohesive-pass-non}. The upper panel shows the data for Cd-centered
clusters while the lower panel shows the data for S-centered clusters. The chosen
scheme of passivation seems to have a pronounced effect for the non-stoichiometric
clusters in the sense apart from very small clusters, the tendency towards formation
in ZB phase seemingly is found to be higher than that in WZ phase in general, irrespective 
of even or odd number of shell, and terminating layer, although in some cases the energy 
difference is indeed tiny (within 1-2 meV) and is within the calculational accuracy. For the non-stoichiometric clusters, study of realistic passivators will be
even more interesting since in many cases the passivating agent itself may have
S/Cd content, giving rise to preferential S or Cd termination of the synthesized clusters.



\subsection{Band Gap Variation with cluster size}

\begin{figure}
\includegraphics[width=7.5cm,keepaspectratio]{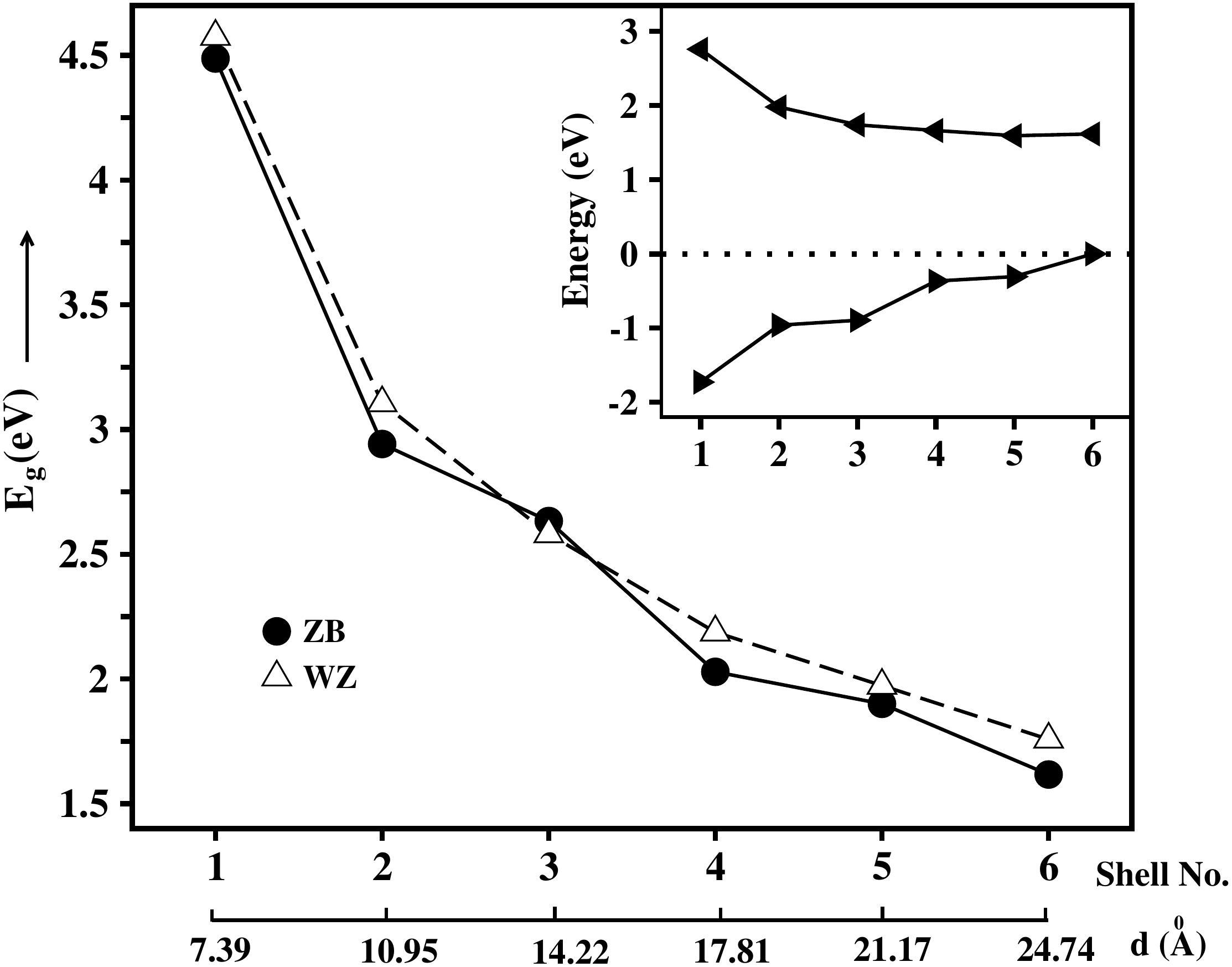}
\caption{Calculated band gaps for stoichiometric clusters plotted as a function of 
increasing cluster size.Inset shows the positions of HOMO (solid right triangles) and
LUMO (solid left triangles) with respect to HOMO of 6th shell cluster, as a function of
 cluster size.}
\label{gap-stoi}
\end{figure}

\begin{figure}
\includegraphics[width=7.5cm,keepaspectratio]{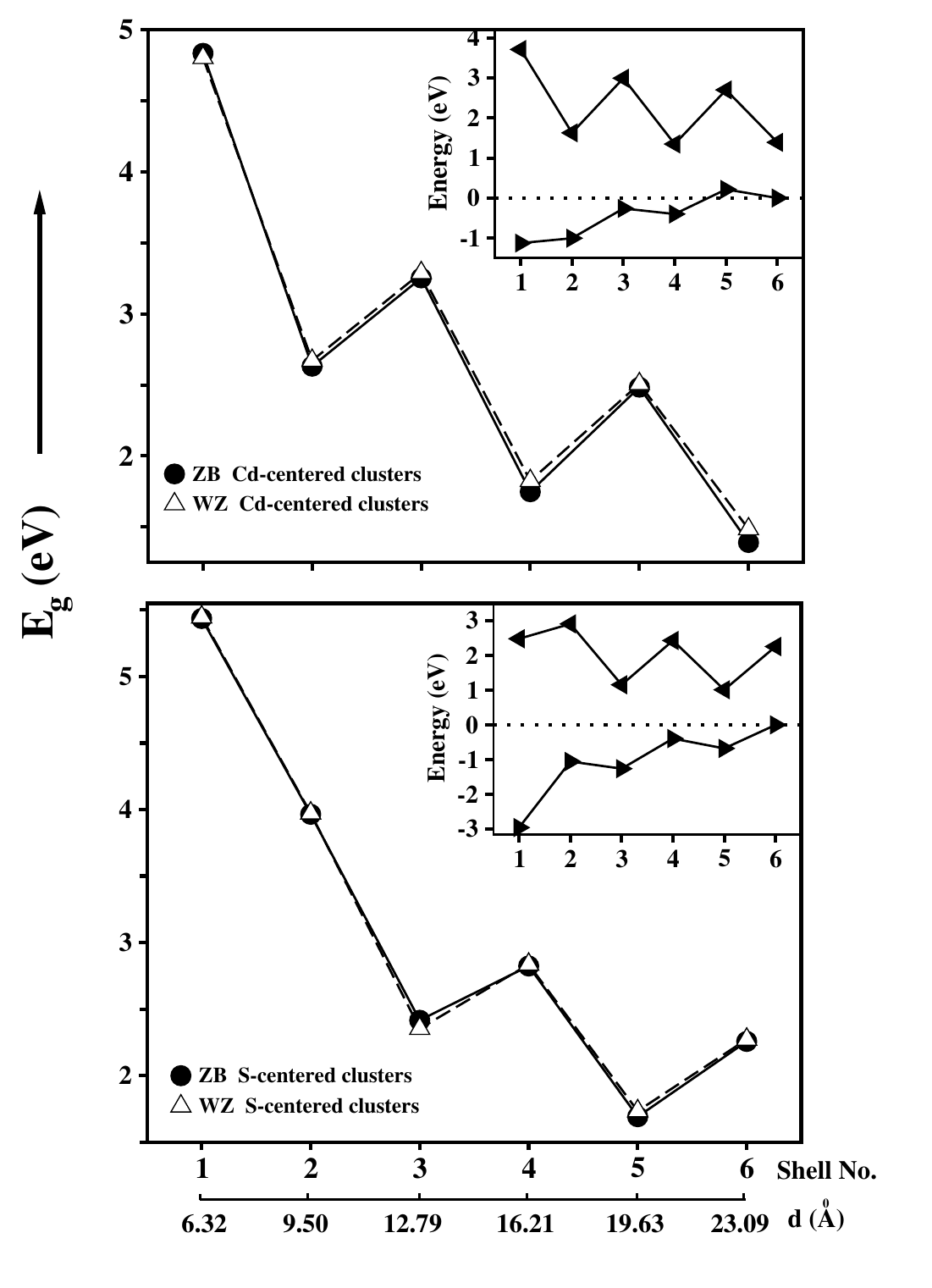}
\caption{Calculated band gaps for non-stoichiometric clusters plotted as a function 
increasing cluster size. Upper and lower panels show the Cd-centered and S-centered
clusters respectively.Inset shows the positions of HOMO (solid right triangles) and
LUMO (solid left triangles) with respect to HOMO of 6th shell cluster, as a function of
 cluster size. }
\label{gap-non}
\end{figure}

The study of CdS clusters in presence of passivator also allows us to investigate
the variation of the band gap as a function of increasing cluster size. In 
Figs.\ref{gap-stoi} and \ref{gap-non}, we show the computed band gaps as a function
of increasing shell size for both ZB and WZ phases, for stoichiometric and
non-stoichiometric clusters. The insets show the individual variations of
the highest occupied molecular orbital (HOMO) and lowest unoccupied molecular
 orbital (LUMO). 
First of all, the band gap associated with WZ structured clusters are found to be
slightly higher than that of the ZB structured clusters in most of the cases,
which results due to the fact that the band gap of the WZ phase in bulk is slightly higher
than that of the ZB phase in bulk.CdS being a direct gap semiconductor, the band gaps of 
WZ and ZB phases are expected to be similar.\cite{Eg_zunger} As expected, the calculated 
band gap shows an overall decrease as a function of increasing cluster size due to
the well known quantum confinement effect, asymptotically approaching the bulk
band gap value in the limit of the infinite cluster size. The calculated band gap for 
the clusters are systematically underestimated due to the overbinding problem related with LDA
treatment of exchange-correlation functional.  For non-stoichiometric
clusters, the band gap variation is found to be highly nonmonotonic. The odd shell 
clusters for Cd-centered clusters and the even shell clusters for S-centered clusters 
show significantly higher value of band gap compared to their respective counterparts.
We therefore conclude that the S-terminated clusters in general show larger band gap
compared to Cd-terminated clusters. As is evident from the variation of HOMO
and LUMO energies shown in the inset, this oscillation is primarily contributed
by the oscillation in the LUMO. The origin of such a behavior lies in the
density of states of the unpassivated cluster itself. Comparing the density
of states of Cd-terminated and S-terminated non-stoichiometric clusters
as shown in left and right panels of Fig.\ref{dos-non}, we found that while
for S-terminated DOS, there exists a well-defined gap in the unoccupied part
of the spectrum, the situation is very different in case of Cd-terminated cluster.
The spectrum is practically gap less or with very small gap in the unoccupied region.
Inclusion of passivating atom, changes the unoccupied spectra drastically in case
of Cd-terminated cluster, while the unoccupied spectra apart from the removal of
the states very close to Fermi energy changes only modestly in case of S-terminated
clusters.

\section{Summary}

Using first-principles density functional based calculations employing plane wave basis set
we present an extensive study of the energy stability and the band gap variation in CdS clusters. 
In particular, we have considered the relative stability between ZB and WZ structures. In order to 
explore the varied experimental conditions, we have considered non-stoichiometric as well as
stoichiometric clusters, in absence and presence of passivating atoms. Our study shows that
the relative stability depends crucially on the surface structure, both geometry and chemistry
depending on the specific cases. This may give rise to highly nonmonotonic behavior of the
relative stability as a function of the growing cluster size. The band gap variation for
the non-stoichiometric clusters is also found to exhibit strong oscillation.



\begin{thebibliography}{99}
\bibitem{note1} In case of WZ structure, due to different values of in-plane and
out-of-plane lattice constants, $a$ and $c$, one of the nearest neighbor has slightly
different distance compared to rest three.
\bibitem{footnote} For a given shell, the Cd to S ratio being same for both the ZB and WZ 
structures, the degree of non-stoichiometry is also same for both the structures. Only for
 3 shell clusters, the Cd to S ratio differs by 3$\%$ between ZB and WZ structures.
\end{thebibliography}
\end{document}